\documentclass[]{article}
\usepackage{amsmath}
\usepackage{amssymb}
\usepackage{graphics}
\usepackage{graphicx}
\usepackage{epsfig}
\usepackage{dcolumn}
\usepackage{bm}
\usepackage{lineno}
\usepackage{multirow}
\usepackage{ragged2e}

\title{\bf A new technique of linseed oil coating in bakelite RPC and the first test results}
\date{}
\begin{document}
		\maketitle
		\vspace*{-1cm}
		\centering
		{
		\author{A.~Sen\footnote[1]{Corresponding Author:
arindam@jcbose.ac.in, arindamsen95@gmail.com},}
		\author{S.~Chatterjee,}
		\author{S.~Das,}
		\author{S.~K.~Ghosh,}
		\author{S.~Biswas}
		}
		\vspace*{0.5cm}
		
		{Department of Physics and Centre for Astroparticle Physics and Space Science (CAPSS), Bose Institute, EN-80, Sector V, Kolkata-700091, India}
		
		\vspace*{0.5cm}
		\centering{\bf Abstract}
		\justify
			Single gap Resistive Plate Chamber (RPC) is one of the very popular gaseous detectors used in high-energy physics experiments nowadays. It is a very fast detector having low cost of fabrication. The RPCs are usually built using glass or bakelite plates having high resistivity $\sim~10^{10}-10^{11}$ $\Omega$~cm. Bakelite RPCs are generally fabricated with a linseed oil coating inside to make the inner electrode surface smoother which helps to reduce the micro discharge probability. Linseed oil coating also reduces the surface UV sensitivity dramatically and effectively protect the bakelite surfaces from the Hydrofluoric Acid (HF), produced by the interaction of fluorine with the water vapour. There is a conventional way to do this linseed oil coating after making the gas gap as done in experiments $e.g.$ ALICE, CMS etc. A new technique is introduced here to do the linseed oil coating on the bakelite plate before making the gas gap. 100\% Tetrafluoroethane (C$_2$H$_2$F$_4$) gas is used to test the RPC module in the avalanche mode with cosmic rays. Conventional NIM electronics is used for this study. The efficiency and noise rate are measured. In this article, the detailed method of fabrication and the first test results are presented. 
			
		\vspace*{0.25cm}

\section{Introduction}

Resistive Plate Chambers (RPCs) are widely used in high-energy physics experiments for their high efficiency, excellent time resolution, hassle-free maintenance and low cost of fabrication \cite{santanico}. In current high-energy physics experiments this parallel plate gaseous detector is mainly used for triggering and tracking purposes due to its excellent time resolution and high efficiency \cite{BABAR, STAR, ALICE, ATLAS, ATLAS2, CMS, HADES}. RPCs are also used in cosmic ray experiments and neutrino experiments for muon detection where large area coverage with minimal cost is required \cite{BESS, ARGO, OPERA, DAYABAY, INO, SB09, INO2}. Future experiments such as Compressed Baryonic Matter 
(CBM) at Facility for Antiproton and Ion Research (FAIR) propose to use RPCs as one of the key detectors for time of flight (TOF) measurements \cite{CBM}.

The RPCs are usually built using glass or bakelite plates having high resistivity $\sim~10^{10}-10^{11}$ $\Omega$~cm. The inner surfaces of the bakelite plates that face the filled gas are usually coated with  linseed oil paint. The linseed oil is a fatty acid (R-COOH), an organic acid that contains the glycerides of linolenic, linoleic, oleic, stearic, and palmitic acids with a high degree of unsaturation of its fatty acid radicals \cite{JV03}. The linseed oil coating reduces the spurious micro discharge on the inner surfaces. If the micro discharge probability is reduced, we can have a better performance compared to the RPC without oil coating \cite{santanico, CL09, sb09_2}. To reduce the after-pulse / noise rate, low UV sensitive material for the electrode is desirable. It is reported earlier that the linseed oil coating  reduces the surface UV sensitivity dramatically \cite{CL09}. Hydrofluoric Acid (HF), produced by the interaction of fluorine with the water vapour, is chemically very reactive. It can affect different materials and has corrosive action. It is reported that the linseed oil coating on the bakelite surface can effectively protect it from the HF vapour attack \cite{CL09}. That means the linseed oil treatment on the inner surfaces of the bakelite electrodes is an essential process for the optimum performance (high efficiency and low noise level) of RPC. For this treatment the inner surfaces of the RPC usually the gas gap is filled with low viscous linseed oil and thinner solution and the liquid is drained out slowly. Dry air is flown through the gas gap to cure the thin linseed oil layer left on all the inner surfaces of the plates as well as those of the spacers \cite{MA97, BH06, 41}. 


However, serious operational problems were observed in the bakelite RPCs in BaBar experiment. It was observed that the conducting paths through the gas gap, mainly around the spacers, created due to the formation of stalagmites by polymerisation of uncured linseed oil droplets, trigger discharges thereby resulting in irreversible damage to the bakelite plates \cite{JV03, 39}. The process of linseed oil treatment was later changed by increasing the ratio of the solvent to produce a thinner coating (10-30 $\mu$m) on the surfaces \cite{79}. Efforts were subsequently made to look for alternatives to linseed oil treatment, or even to develop bakelite sheets that can be used without the application of linseed oil  \cite{41}.
	
One of the main limitations of RPC is that its low particle rate handling capability \cite{YH13}. Nowadays RPC detector can handle a particle rate $\sim$~10 kHz/cm$^{2}$ in the avalanche mode of operation \cite{LP20} but for the future experiments, detectors with high particle rate handling capability ($\sim$~15 kHz/cm$^{2}$) is required \cite{EN}. One of the ways of increasing the particle rate handling capability in RPC is the use of low resistive electrode plates of the detector. It is to be mentioned also that the mode of operation depends on used gas and applied voltage \cite{meghna}. For high rate operation, the option is the avalanche mode. 

Keeping in mind the possibility of using bakelite RPCs as future high rate capable tracking detectors, we have taken up a study to characterise RPC prototypes built using a particular type of bakelite plates with moderate bulk resistivity. The first prototype was built without any oil coating inside, with indigenous bakelite plates having bulk resistivity $3~\times~10^{10}$ $\Omega$~cm. The prototype was tested with 100\% Tetrafluoroethane (C$_2$H$_2$F$_4$) gas. With this prototype, an efficiency $\sim$~70\% was obtained with an applied voltage of 10.2~kV onwards \cite{asen}.

In the present work, we built a linseed oil coated bakelite RPC. However, in this work, we have adopted a different technique for the linseed oil treatment. In contrary to the usual procedure, the bakelite plates are coated with linseed oil before making the gas gap. 

The prototype is tested with cosmic ray using 100~\% Tetrafluoroethane (C$_2$H$_2$F$_4$) gas and conventional NIM electronics. In this article, we report the first result of the bakelite RPC, fabricated using a new technique of linseed oil coating.

\section{Fabrication of the detector}

\begin{figure}[htb!]
		\centering{
			\includegraphics[scale=0.45]{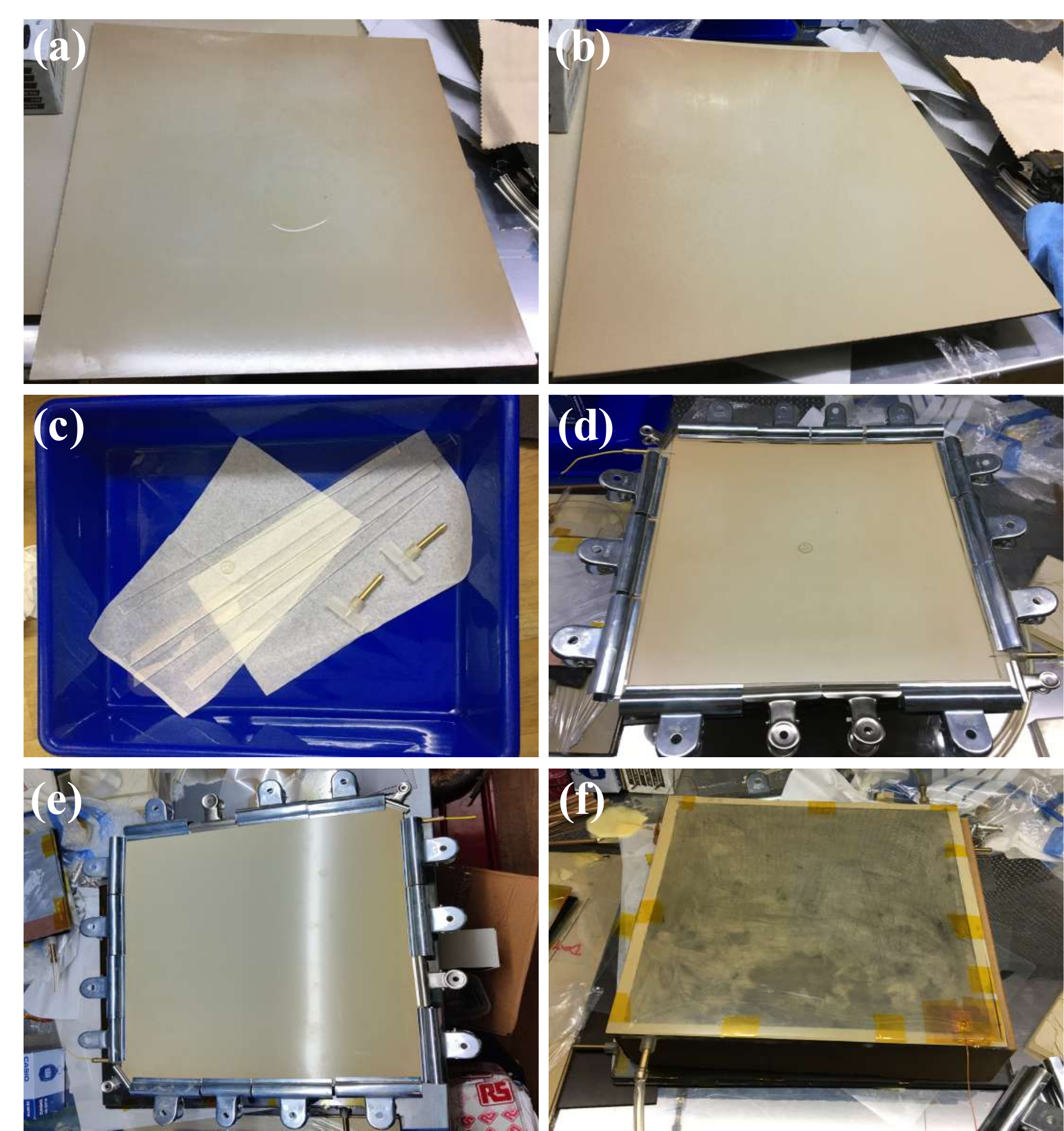}
		}
		\caption{Steps of building the RPC module. (a) Application of linseed oil on the bakelite surface, (b) cured linseed oil coated bakelite surface, (c) polycarbonate made gas nozzles and spacers, (d) gluing of spacers and nozzles on one bakelite plate, (e) making of gas gap after gluing the second plate, (f) complete RPC module after graphite coating.}\label{build}
	\end{figure}

Two bakelite plates having a dimension 27~cm~$\times$~27~cm and thickness 2~mm are used as the electrodes. The bulk resistivity of the plates is measured to be $\sim$ 3 $\times$ 10$^{10} \Omega$~cm at 22$^\circ$C temperature and 60\% relative humidity. At first, the bakelite plates are cleaned using an isopropyl alcohol bath. Commercially available linseed oil is used for the inside coating. The linseed oil coating is done here in a similar way as the silicone oil coating technique followed in Ref.~\cite{SB09}. About 2~g of linseed oil is applied over the 27~cm~$\times$~27~cm area of each plate. Based on the specific gravity (0.930 at 15.5~$^\circ$C) of the fluid, the estimated coating thickness would be $\sim$~30~$\mu$m. The linseed oil is distributed over the surfaces and both the plates are left for 15 days in a sealed box for curing. Two plates are then cleaned again with dry air. Uniform separation between the electrode plates is ensured by using four edge spacers of dimension 27~cm~$~\times$~1~cm and thickness 2~mm and one button spacer of diameter 10~mm and thickness 2~mm. All the spacers are made of polycarbonate. Two nozzles also made of polycarbonate (resistivity $\sim$ 10$^{15}$ $\Omega$~cm) for gas inlet and outlet are used as part of edge spacers. All the spacers are glued on one plate in the oil coated side using Araldite epoxy adhesive. The glued plates are kept for 24 hours for curing and the other plate is glued on it and kept again for curing for 24 hours. All the gluing processes are carried out on a laminar flow table. After cleaning, a thin layer of graphite is coated at the outer surfaces of the bakelite plates for the distribution of voltage. A gap of 1~cm from the edges of the plate to the graphite layer is maintained to avoid external electrical discharge. The average surface resistivity of the two graphite layers are found to be $\sim$~510~k$\Omega$/$\Box$ and 540~k$\Omega$/$\Box$  respectively. Two 1~cm~$\times$~1~cm copper tapes of $\sim$~20~$\mu$m thickness are pasted at two diagonally opposite corners to apply high voltage (HV). The HV cables are soldered on these copper strips. These copper strips are covered using Kapton tapes for isolation. Equal HVs with opposite polarities are applied on two surfaces. The steps of building of the module are shown in Figure~\ref{build}.

2.5~cm wide copper (20~$\mu$m thick) pick-up strips are fabricated having 2~mm separation among two consecutive strips, in order to collect the accumulated induced charge and are placed above the graphite layers. The pick-up strips are made by the etching process of one side from a double sided copper cladded, 2~mm thick G-10 sheet. The other side copper layer of the G-10 sheet is used as the ground plane. The strips are covered with 100~$\mu$m thick mylar foils to isolate them from the graphite layers. The signals from the strips are collected through RG-174/U coaxial cables.
	
\section{Experimental set-up}

Three plastic scintillation detectors, two placed above the RPC module and one placed below, are used to obtain the trigger from the incoming cosmic rays. The coincidence signal obtained from the top most paddle scintillator (SC1) having dimension 10~cm~$\times$~10~cm, a finger scintillator (SC2) of dimension 10~cm~$\times$~2~cm and the paddle scintillator (SC3) having dimension 20 cm $\times$ 20 cm are taken as the trigger (3-fold). RPC is placed in between the finger (SC2) and the paddle scintillator (SC3). All the scintillators are operated at 1550~V and -15~mV threshold is applied to the leading edge discriminator (LED). 

\begin{figure}[htb!]
		\centering{
			\includegraphics[scale=0.30]{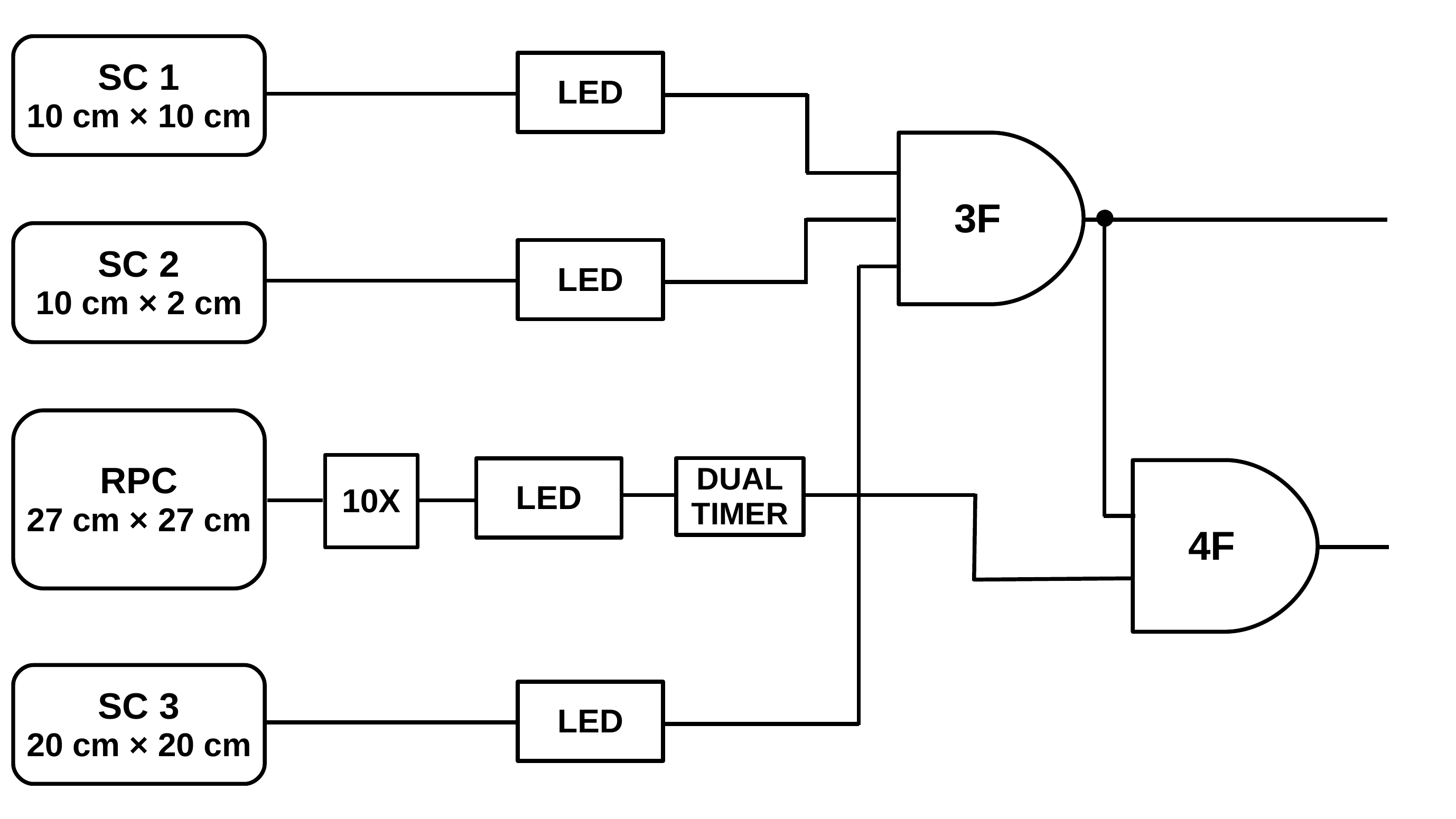}
		}
		\caption{Schematic of the experimental set-up}\label{ckt}
	\end{figure}

The RPC signal from the pick-up strip is first fed to the 10x fast amplifier and the output of it then goes to the LED. Different thresholds are applied to the LEDs to reduce the noise. From the LED one output goes to the scalar to count the number of the signal from the RPC which is known as the noise count or singles count of the chamber. The other output from LED goes to the dual timer where the discriminated RPC signal is stretched to avoid any double counting of the pulses and also to apply the proper delay to match the signal with the trigger. The output of the dual timer is put in coincidence with the trigger and this is defined as the 4-fold. The window of the cosmic ray test set-up is of area 10~cm~$\times$~2~cm. The detailed block diagram of this arrangement is shown in Figure~\ref{ckt}.

	
100\% Tetrafluoroethane (C$_2$H$_2$F$_4$) gas is used as the active medium. A typical gas flow rate of 2~ml/min equivalent to 6 gap volume changes per day is maintained by using two needle valves.

The HVs to the RPC module are applied at a ramp-up rate of 2~V/s on both the sides. The leakage currents from both the side as measured by the HV module are also recorded. The temperature and the humidity are also recorded at the time of measurement using a data logger, built-in house \cite{sahu}.

\section{Result}

\begin{figure}[htb!]
		\centering{
			\includegraphics[scale=0.45]{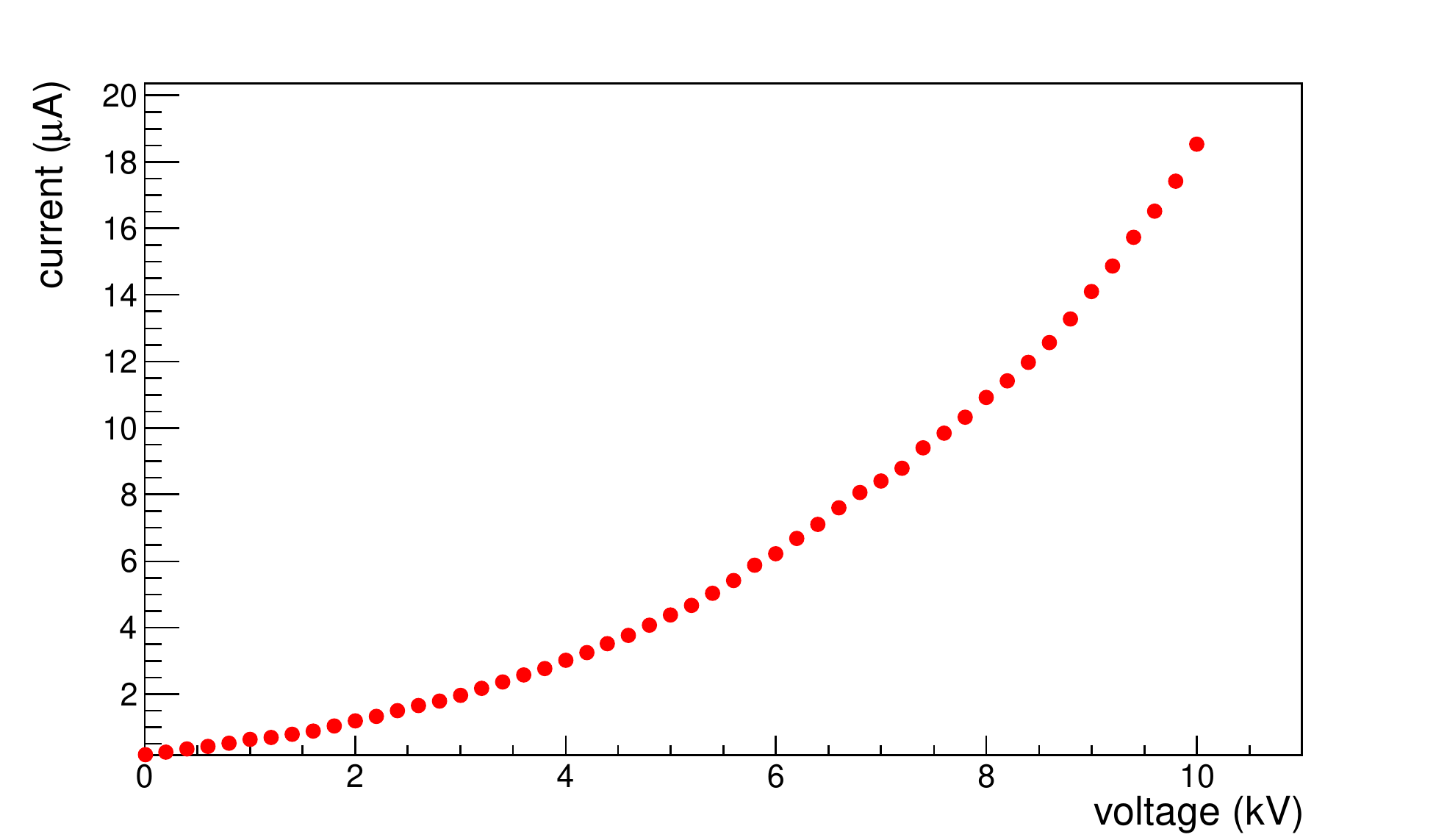}
		}
		\caption{Leakage current as a function of the applied voltage for the RPC module.}\label{iv}
	\end{figure}

After building the chamber, 100\% C$_2$H$_2$F$_4$ gas is purged for 24 hours before application of HV. To check the performance of the detector, firstly the leakage current through the RPC module is measured as a function of the applied HV and shown in Figure~\ref{iv}. Breakdown of the gas, although not sharp, is seen at about $\sim$~8~kV. The gas gap behaves as an insulator in the low applied voltage range and hence the slope over this voltage region scales as the conductance of the polycarbonate spacers. At higher range of voltage, the gas behaves as a conducting medium due to the ionisation. Therefore, the slope over this range scales as the conductance of the bakelite plates. The curve of leakage current vs. voltage as found here is not only very similar to the curve as reported earlier \cite{MA97} but the magnitude of current is also comparable.  

	
	\begin{figure}[htb!]
		\centering{
			\includegraphics[scale=0.45]{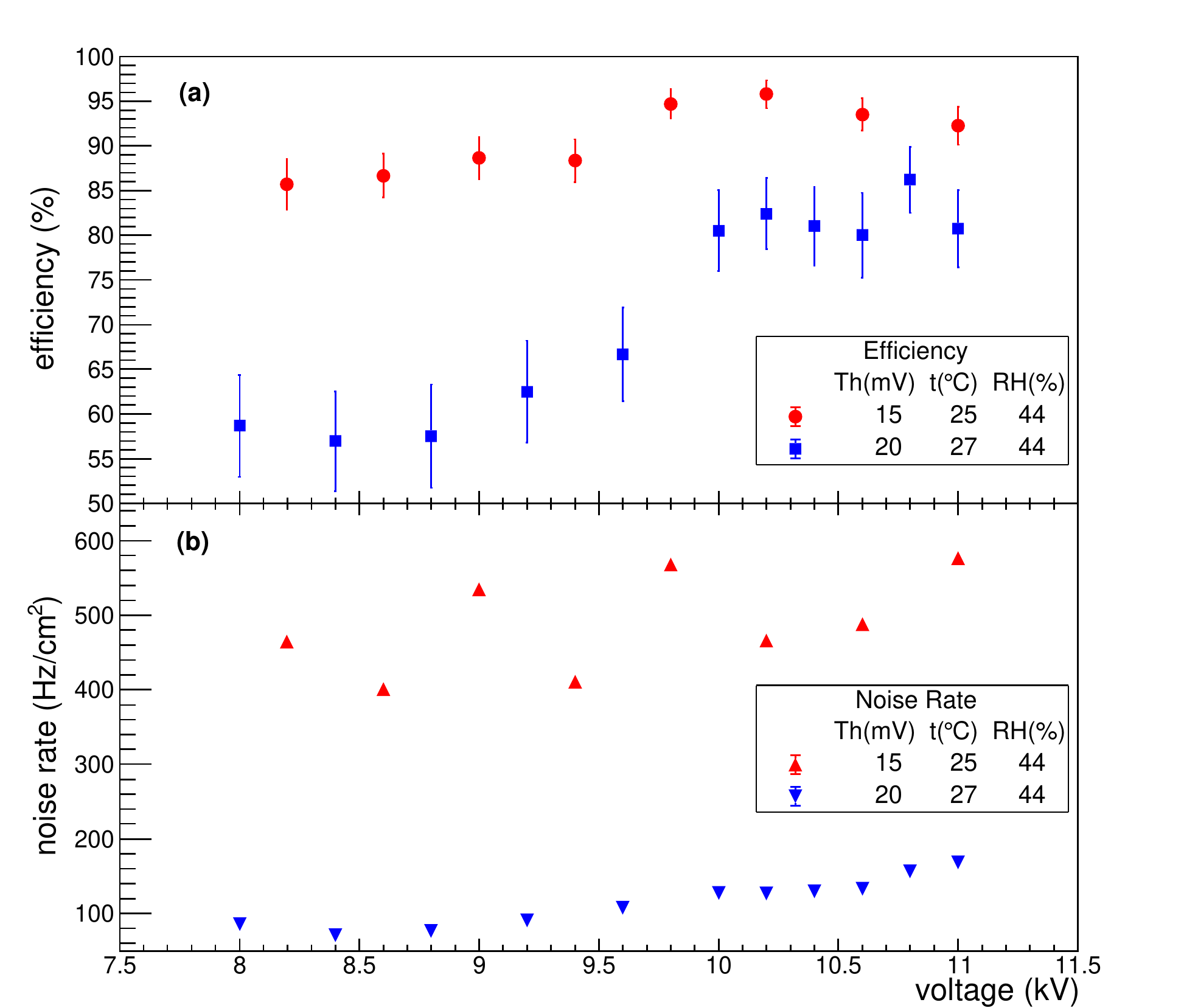}
		}
		\caption{(a) The efficiency as a function of the applied voltage for the RPC, (b) Noise rate as a function of the applied voltage.}\label{noise}
	\end{figure}


The efficiency of the RPC module for the cosmic rays, defined as the ratio of the 4-fold counts of the RPC and the 3-fold trigger count of the plastic scintillator telescope for a fixed duration and the noise rate (or singles count rate) of the RPC, defined as the number of counts per unit area of the strip per second, are studied by varying the applied HV. Both the efficiency and noise rate are measured for two different discriminator threshold settings, -15~mV and -20~mV respectively for the RPC. The temperature and relative humidity values during these measurements are recorded and the average temperature is found to be about 25 $^\circ$C and 27 $^\circ$C, respectively and the average relative humidity is found to be 44\% in both the cases. The efficiency and the noise rate as a function of voltage is shown in Figure~\ref{noise}. For -15~mV threshold setting the efficiency increases with applied voltage and reaches a plateau at $\sim$~95\% from 9.4~kV onwards whereas for -20~mV threshold setting the efficiency saturates at $\sim$~85\% from the applied voltage of 10.1~kV onwards. The noise rate increases with applied HV. The noise rate is measured to be much higher for the lower threshold with a maximum value of $\sim$~500 Hz/cm$^{2}$. For -20~mV threshold the maximum noise rate is found to be $\sim$~200 Hz/cm$^{2}$.

\section{Summary and Outlook}
	
    A small size RPC prototype is built having a dimension of 27~cm~$\times$~27~cm bakelite plates of thickness 2~mm. The gas gap of the prototype is also made 2~mm. A new technique is followed for the  linseed oil coating of the bakelite sheets. In conventional bakelite RPC, the linseed oil coating is done after making the gas gap. In case of BaBar RPC the performance deteriorated drastically due to trapping of uncured linseed oil in the hidden storage cavities near the spacers \cite{JV03}. To ensure that the curing is properly done throughout the entire surface, in this particular work, the linseed oil coating is done before making the gas gap. After the linseed oil coating, the plates are cured for 15 days. The advantage of this procedure is that after linseed oil coating it can be checked visually whether the curing is properly done or any uncured droplet of linseed oil is present or not. 

The detector is tested with 100\% Tetrafluoroethane (C$_2$H$_2$F$_4$) gas in the avalanche mode. Efficiency plateau $\sim$~95\% from 9.4~kV onwards and $\sim$~85\% from 10.1~kV onwards are obtained for the -15~mV and -20~mV discriminator threshold respectively. The noise rate is found to be very high for such a chamber compared to the results reported earlier for the RPCs with bakelite plates oil coated in the conventional way \cite{bergnoli1, bergnoli2, bossu}.

The detection efficiency of the RPC in the cosmic ray test is compared with the conventional linseed oil coated bakelite RPC \cite{santanico, MA97, BH06}. However, the investigation behind the high leakage current and high noise rate is going on. The long-term stability test, measurement of timing properties and the effect of the temperature and relative humidity on the performance are planned to be carried out in future.

\section*{Acknowledgement}
	
We would like to thank Mr. Shivshant Chauhan for his help in the installation of the detector and running the experiment. We would also like to thank Mr. Subrata Das for helping in fabrication of the pick-up strips used in this study. The authors would also like to thank Prof. Sibaji Raha, Prof. Rajarshi Ray and Dr. Sidharth K. Prasad for valuable discussions and suggestions in the course of the study. This work is partially supported by the CBM-MuCh project from BI-IFCC, DST, Govt. of India. A. Sen acknowledges his Inspire Fellowship research grant [DST/INSPIRE Fellowship/2018/IF180361].


\noindent
	
\begin{thebibliography}{99} 
		
\bibitem{santanico} R. Santonico and R. Cardarelli, Nucl. Inst. and Meth. A {\bf187} (1981) 377.

\bibitem{BABAR} BaBar, Technical Design Report, SLAC-R-95-457.

\bibitem{STAR} E. Cerron Zeballos {\it et al.}, Nucl. Instr. and Meth. A {\bf374} (1996) 132.

\bibitem{ALICE} R. Arnaldi {\it et al.}, Nucl. Instr. and Meth. A {\bf451} (2000) 462.

\bibitem{ATLAS} ATLAS Collaboration, ATLAS Muon Spectrometer Technical Design Report, CERN/LHCC, Vol. 97-22.

\bibitem{ATLAS2} G. Chiodini {\it et al.}, Nucl. Inst. and Meth. A {\bf581} (2007) 213.

\bibitem{CMS} CMS Collaboration, Muon Project, CERN/LHCC 97-32.

\bibitem{HADES} D. Belver {\it et al.}, Nucl. Instr. and Meth. A {\bf602} (2009) 687.

\bibitem{BESS} J. Han {\it et al.}, Nucl. Instr. and Meth. A {\bf577} (2007) 552.

\bibitem{ARGO} G. Aielli {\it et al.}, Nucl. Instr. and Meth. A {\bf562} (2006) 92.

\bibitem{OPERA} R. Acquafredda {\it et al.}, JINST {\bf4} (2009) P04018.    

\bibitem{DAYABAY} Q. Zhang {\it et al.}, Nucl. Instr. and Meth. A {\bf583} (2007) 278.

\bibitem{INO} http://www.ino.tifr.res.in/ino/

\bibitem{SB09} S. Biswas {\it et al.}, Nucl. Instr. and Meth. A {\bf602} (2009) 749.

\bibitem{INO2} A. Behera {\it et al.}, Nucl. Instr. and Meth. A {\bf602} (2009) 784.

\bibitem{CBM} I. Deppner {\it et al.}, Nucl. Instr. and Meth. A {\bf 661} (2012) S121.

\bibitem{JV03} J. Va'vra, Nucl. Instr. and Meth. A {\bf515} (2003) 354.

\bibitem{CL09} C. Lu, Nucl. Instr. and Meth. A {\bf602} (2009) 761.

\bibitem{sb09_2} S. Biswas {\it et al.}, Nucl. Instr. and Meth. A {\bf604} (2009) 310.	

\bibitem{MA97} M. Abbrescia {\it et al.}, Nucl. Instr. and Meth. A {\bf394} (1997) 13.

\bibitem{BH06} B. Hong {\it et al.}, Journal of the Korean Physical Society, Vol. {\bf48}, No. 4, April 2006, 515.

\bibitem{41} J. Zhang {\it et al.}, Nucl. Instr. and Meth. A {\bf540} (2005) 102.

\bibitem{39} F. Anulli, {\it et al.}, Nucl. Instr. and Meth. A {\bf508} (2003) 128.

\bibitem{79} F. Anulli, {\it et al.}, Nucl. Instr. and Meth. A {\bf539} (2005) 155.

\bibitem{YH13} Y. Haddad {\it et al.}, Nucl. Instr. and Meth. A {\bf718} (2013) 424.

\bibitem{LP20} L. Pizzimento {\it et al.}, JINST {\bf15} (2020) C11010.

\bibitem{EN} E. Nandy {\it et al.}, Proc. of the DAE-BRNS Symp. on Nucl. Phys. {\bf61} (2016) 1024.

\bibitem{meghna} K. K. Meghna {\it et al.}, Nucl. Instr. and Meth. A {\bf816} (2016) 1.

\bibitem{asen} A. Sen {\it et al.}, JINST {\bf15} (2020) C06055.



\bibitem{sahu} S. Sahu, {\it et al.}, JINST {\bf12} (2017) C05006.

\bibitem{bergnoli1} A. Bergnoli {\it et al.}, Nuclear Physics B (Proc. Suppl.) {\bf158} (2006) 93.

\bibitem{bergnoli2} A. Bergnoli {\it et al.}, IEEE TRANSACTIONS ON NUCLEAR SCIENCE VOL. {\bf52}, NO. 6, DECEMBER 2005.

\bibitem{bossu} F. Bossu {\it et al.}, JINST {\bf7} (2012) T12002.


		
	\end{thebibliography}
\end{document}